\documentclass[a4paper]{spie}

\usepackage{amsmath,amsfonts,amssymb}
\usepackage{graphicx}
\usepackage[colorlinks=true, allcolors=blue]{hyperref}
\usepackage{siunitx}
\usepackage[separate-uncertainty=true, multi-part-units=single]{siunitx}
\DeclareSIUnit\arcmin{arcmin}

\title{Dish Assembly Precision for HIRAX}

\author[a]{Jennifer Studer}
\author[a]{Devin Crichton}
\author[a]{Alexandre Refregier}
\author[b, c]{Keshav Bechoo}
\author[d]{H. Cynthia Chiang}
\author[b]{Chandrachud B. V. Dash}
\author[b]{Sindhu Gaddam}
\author[d, e]{Aditya Krishna Karigiri Madhusudhan}
\author[g]{Martin Kunz}
\author[b, c]{Kavilan Moodley}
\author[b, c]{Warren Naidoo}
\author[b, c]{Tasmiya Papiah}
\author[b, h]{Isibabale Qhoboshiyane}
\author[b]{Liantsoa F. Randrianjanahary}
\author[f]{Benjamin R. B. Saliwanchik}
\author[g]{Ajith Sampath}
\author[b, c]{Corrie Ungerer}
\author[a]{Thierry Viant}
\author[b, c]{Anthony Walters}

\affil[a]{Institute for Particle Physics and Astrophysics, ETH Zurich, Zurich, Switzerland}
\affil[b]{Astrophysics Research Centre, University of KwaZulu-Natal, Westville Campus, Durban, 4000, South Africa}
\affil[c]{School of Agriculture and Science, University of KwaZulu-Natal, Westville Campus, Durban, 4000, South Africa}
\affil[d]{Department of Physics, McGill University, 3600 Rue University, Montreal, QC H3A 2T8,
Canada}
\affil[e]{Trottier Space Institute, McGill University, 3550 University Street, Montreal, QC H3A 2A7, Canada}
\affil[f]{Instrumentation Department, Brookhaven National Laboratory, Upton, NY, USA}
\affil[g]{Department of Theoretical Physics, University of Geneva, 24 quai Ernest-Ansermet, CH1211 Genève 4, Switzerland}
\affil[h]{University of Zululand Department of Mathematical Sciences, KwaDlangezwa, 3886, South Africa}

\authorinfo{Further author information: (Send correspondence to J. Studer: E-mail: studerje@ethz.ch)}

\begin{document} 

\maketitle

\begin{abstract}
The Hydrogen Intensity and Real-time Analysis eXperiment (HIRAX) is a radio interferometer array that is being deployed at the South African Radio Astronomy Observatory (SARAO) Square Kilometer Array (SKA) site in South Africa. Mapping the southern sky, its aim is to observe neutral hydrogen (HI) through intensity mapping (IM) across the redshift range of $0.78$-$2.55$. The observation of HI makes it possible to tomographically probe large cosmological volumes, enabling constraints on, for example, the dark energy equation of state. Systematics are a significant concern in deriving cosmological constraints from HI IM due to the presence of strong foreground signals. Instrumental effects such as dish surface deviations and feed placement errors need to be carefully controlled and monitored to preserve the sensitivity of HIRAX's redundant array configuration. These instrumental systematics cause bright foreground power to leak into the faint cosmological signal. The 6 m parabolic dishes are made from fiberglass with an embedded aluminum mesh that acts as the reflector. The feed is held in place at the prime focus ($f/D = 0.21$) by four fiberglass legs. This paper presents the dish surface and feed placement precision of the first 28 dishes, derived from photogrammetry metrology measurements taken during the dish fabrication, as well as from measurements in the field. While surface deviations are predominantly well within requirements, feed placement satisfies all accuracy criteria but fails precision limits; however, precision is expected to improve with a larger sample size. These results establish a baseline for modeling primary beam effects, ultimately enabling the characterization and mitigation of systematic errors in the HI IM signal with the full HIRAX array.
\end{abstract}


\section{Introduction} \label{sec:intro} 
The Hydrogen Intensity and Real-time Analysis eXperiment (HIRAX) is a new radio interferometer, being built in the Karoo at the South African Radio Astronomy Observatory (SARAO) Square Kilometer Array (SKA) site. Its primary goal is to map the large-scale structure of the universe by observing the 21 cm neutral hydrogen (HI) line via intensity mapping (IM) in the redshift range of $0.78$-$2.55$, corresponding to a frequency range of \qtyrange[range-units=single,range-phrase=-]{400}{800}{\mega\hertz}. The 21 cm spin flip is a forbidden transition~\cite{Liu_2020} that we can only observe due to the high abundance of HI. Using intensity mapping, which maps the total received intensity without resolving individual sources, HIRAX aims to map one-third of the southern sky over a four-year period. Observing this large cosmological volume will enable a sensitive probe of the geometric expansion of the universe and, in particular, constraints on the behavior and evolution of dark energy.
\begin{figure}[ht]
    \centering
    \includegraphics[width=0.75\textwidth]{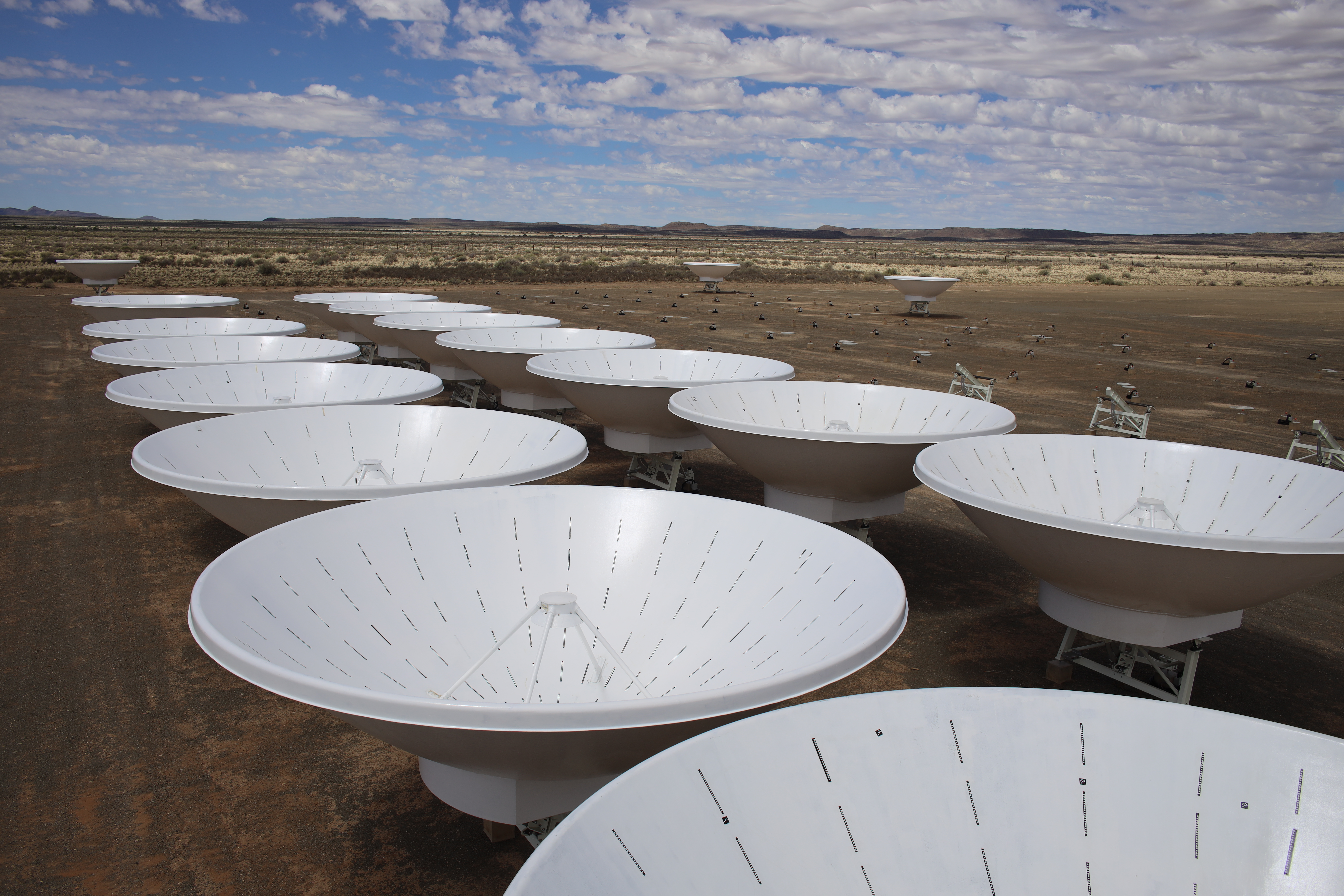}
    \caption{The HIRAX array during commissioning in February 2026 at the Swartfontein site, South Africa.}
    \label{fig:hirax}
\end{figure}
This effort  complements the Canadian Hydrogen Observatory and Radio-transient Detector (CHORD)~\cite{CHORD}{}, which will conduct a similar survey in the northern hemisphere and whose dish design is shared with HIRAX. Both telescopes are radio interferometers with a compact, redundant array layout to increase sensitivity to the cosmological scales of interest (see Figure \ref{fig:hirax} for the HIRAX array under construction in February 2026). Other telescopes with the goal of measuring the cosmological 21 cm signal include HERA~\cite{DeBoer_2017}{}, CHIME~\cite{Amiri2022}{}, Tianlai~\cite{Wu2020}{}, LOFAR~\cite{VanHaarlem2013}{}, NenuFAR~\cite{Munshi2025}{}, BINGO~\cite{Abdalla2021}{}, PUMA~\cite{Ahmed2019}, ASKAP~\cite{Hotan2021}, MeerKAT~\cite{Jonas2016}, SKA Mid~\cite{Swart2022}{}, and SKA Low~\cite{Labate2022}{}.\\
However, extracting the faint cosmological 21 cm signal presents a significant hurdle, as astrophysical foregrounds are up to six orders of magnitude brighter~\cite{Crichton_2022}{}. Instrumental systematic errors can distort the signal, causing foreground power to leak into the clean cosmological window~\cite{Morales2018}{}. Therefore, it is essential to mitigate and understand these effects thoroughly. A primary source of these distortions arises during beam calibration~\cite{Kim_2022, Sampath2026}{} from the inherent properties of the instrument. The telescope's directional gain response, known as the primary beam, is given by the instrument configuration. Specifically, dish surface residuals and feed positioning errors deform the beam, and need to be kept within tight tolerances.\\
In this paper, we employ metrology~\cite{Mohammad2022}{} to measure and verify the dish surface and feed positioning. While a combination of photogrammetry and laser tracking is used and described in~\cite{MetrologyST}{}, this work only utilizes photogrammetry for data acquisition. The photogrammetry system used is a  DPA (Digital Photogrammetric Analysis) Professional system\footnote{\url{https://hexagon.com/products/dpa-professional}} and the accompanying AICON 3D Studio software. Its main advantage is that it allows rapid, repeatable measurements with a consistent point density and distribution, making it significantly faster than laser tracker measurements. Furthermore, the reflectometer measurements presented in \cite{MetrologyST} showed that the standard deviation of the distance between the dish surface and the embedded reflective mesh is $0.1$ mm. This demonstrates that it is sufficient to measure the surface of the fiberglass dish alone instead of measuring the reflective aluminum mesh embedded within the dish.\\
The paper is organized as follows.  Section \ref{sec:requirements} outlines the accuracy and precision requirements for both dish surface and feed positioning. Section \ref{sec:surface} presents the surface measurements and evaluates their resulting accuracy and precision using measurements taken in the factory and on-site. Finally, the placement of the feed within the dish is examined in Section \ref{sec:feed}.

\section{Requirements} \label{sec:requirements}
\begin{figure}[ht]
    \centering
    \includegraphics[trim=0 150 180 40, clip, width=1.0\textwidth]{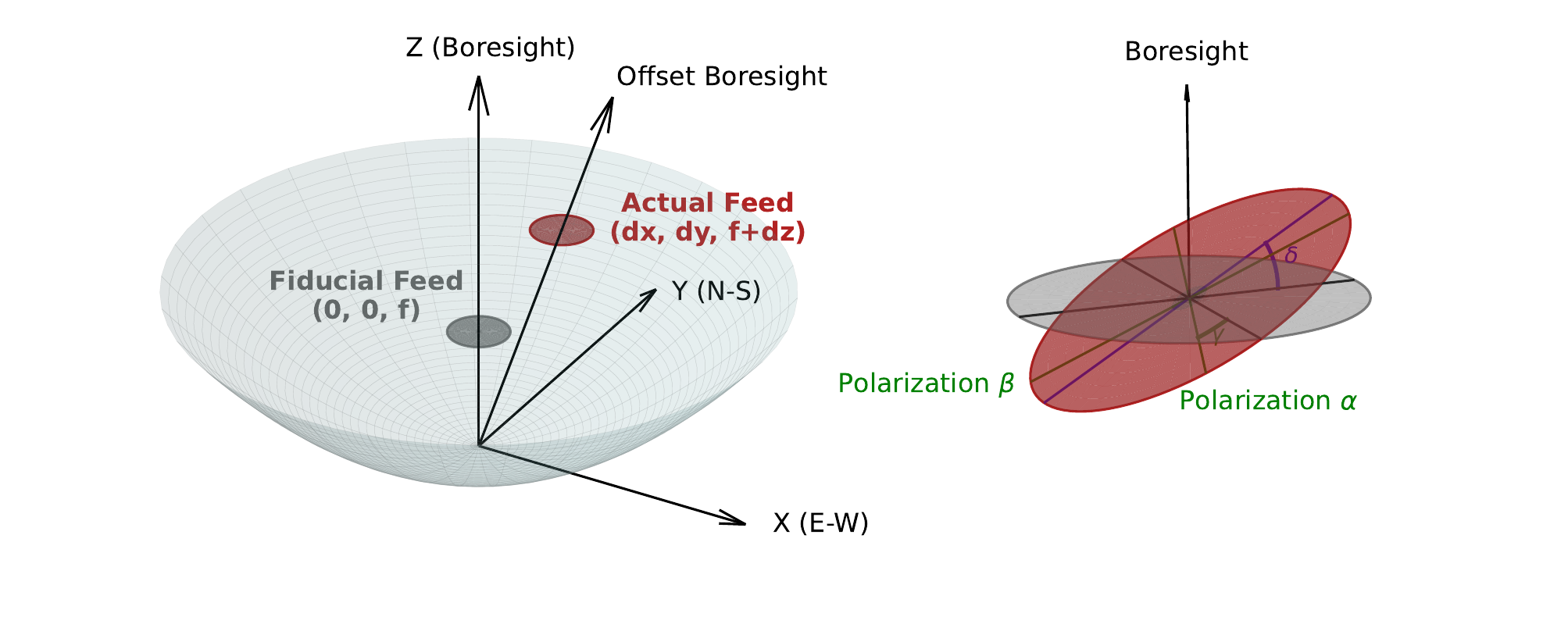}
    \caption{Illustration showing the actual feed position described by translational and rotational parameters relative to the fiducial feed position at the best-fit focal point of the parabolic dish. The translational offset are given by $\mathrm{d}x$, $\mathrm{d}y$ and $\mathrm{d}z$, while $\delta$ and $\gamma$ represent the feed inclination and the alignment offset from the dishes N-S axis, respectively.}
    \label{fig:dishcoord}
\end{figure}

In order to reach the sensitivity required to observe the cosmological 21 cm signal and to perform redundant array calibration, which assumes all the primary beam responses across the array to be nearly identical, strict accuracy and precision requirements have been put in place that must be realized in the dish production and assembly. These requirements are presented in detail in the HIRAX Telescope Mechanical Assembly Requirements Document\footnote{\url{https://hirax.ukzn.ac.za/wp-content/uploads/2021/12/HIRAX_REQ001_002_V1_Baselined-signed.pdf}}. In this paper, we focus on the requirements related to the placement of the feed relative to the focal position and the dish surface.\\
For this analysis, we define a dish-oriented coordinate system as shown in Figure \ref{fig:dishcoord}, which is determined by the best-fit paraboloid of each dish. Here, the vertex of the dish defines the origin, the boresight vector defines the $z$-axis, and the $x$-axis is aligned with the East-West orientation of the dish. Ideally, the fiducial feed position is located at the focal point of the parabolic dish. 
As illustrated in Figure \ref{fig:dishcoord} the actual position typically differs by an offset described by $(\mathrm{d}x, \mathrm{d}y, f+\mathrm{d}z)$, where $f$ is the best-fit focal length and $\mathrm{d}x$, $\mathrm{d}y$ and $\mathrm{d}z$ are the deviations from the fiducial feed position in the $x$, $y$ and $z$ direction, respectively. The feed can also be tilted by an angle $\delta$ with respect to the $xy$-plane in which it is intended to lie. The offset direction of the inclination angle $\delta$ is also recorded for potential use in future analysis, but it is not part of the requirements and therefore it is not further discussed in this paper. Additionally, all feeds $\alpha$ polarization axes must align with the N-S axis of the dish to ensure consistent polarization alignment on-site. We use the angle $\gamma$ to describe deviations from the ideal N-S alignment, where $\gamma$ lies in the tilted plane.\\
The accuracy and precision requirements for both the feed parameters and the dish surface are shown in Table \ref{tab:TMA_req} and further described below. 
\begin{table}[t]
\centering
\begin{tabular}{|c|c|c|c|} 
\hline
Telescope Mechanical Assembly Parameters & Accuracy & Target Precision & Units \\
\hline
Feed position $\mathrm{d}x$, $\mathrm{d}y$ and $\mathrm{d}z$ & 3.0 & 0.5 & mm\\
Feed orientation angle $\delta$ & 5.0 & 2.5 & arcmin \\
N-S alignment $\gamma$ & 7.5 & 2.5 & arcmin\\
Dish Surface & 5.0 & 1.0 & mm\\
\hline
\end{tabular}
\caption{Accuracy and precision requirements for the telescope mechanical assembly parameters as described in Section \ref{sec:requirements}. The accuracy for the feed parameters is given by Equation \ref{eq:mean} and the precision is computed according to Equations \ref{eq:prec} and \ref{eq:surfprec}, respectively.}
\label{tab:TMA_req}
\end{table}
The HIRAX array utilizes a highly redundant layout to maximize sensitivity and enable redundant calibration. Redundant array calibration relies on the assumption that telescope pairs with identical baselines observe identical sky signals. This assumption demands stringent tolerances on both the accuracy and precision of the telescope's mechanical parameters. Otherwise, averaging signals that possess slight physical variations introduces calibration artifacts.\\  
The accuracy of the single parameters\footnote{The definitions and equations for the accuracy and precision of both the single parameters and the dish surface are adopted from the HIRAX Telescope Mechanical Assembly Requirements Document to ensure a consistent comparison against the required thresholds. Consequently, we also adopt the statistical conventions of that document, which utilize $1/N$ rather than the Bessel-corrected $1/(N-1)$ for population metrics.} (which encompass all feed-related metrics) is defined as the absolute difference between the mean of the measured values $\overline{p}$ and the ideal value, where the mean value of $N$ number of dishes is given by 
\begin{equation} \label{eq:mean}
    \overline{p} = \frac{1}{N}\sum_{i=1}^N p_i.
\end{equation} 
The precision for these single parameters is defined as the root-mean-square (RMS) error from the mean: 
\begin{equation} \label{eq:prec}
    \sigma_{p} = \sqrt{ \frac{1}{N}\sum_{i=1}^N (p_i - \overline{p})^2 }.
\end{equation}
The accuracy and precision for the dish surface must be calculated differently, as surface measurements of a single dish yield a three-dimensional point cloud rather than a single scalar value. For each dish, a paraboloid of revolution $p_i(x,y)$ is fitted to the dataset using a least-squares approach as described in~\cite{MetrologyST}{}. Similar to the single-parameter method, we define a mean best-fit paraboloid $\overline{p}(x,y)$ from the measurement of all $N$ dishes. The surface accuracy is then defined as the maximum deviation of this mean best-fit paraboloid $\overline{p}(x,y)$ from the fiducial paraboloid $p_0(x,y)=\frac{1}{4f_0}(x^2+y^2)$, which has a focal length of $f_0=1260$ mm. The surface precision is computed for each dish $\sigma_i(x,y)$ and given by
\begin{equation} \label{eq:surfprec}
    \sigma_i(x,y) = \sqrt{ \frac{1}{N_p}\sum_{j=1}^{N_p} (P_{ij} - \overline{p}(x,y))^2 },
\end{equation}
where $P_{ij}$ is the measured point cloud interpolated onto a polar grid for dish $i$ and $N_p$ the total number of measurement points.\\
Because evaluating the accuracy and precision requires a statistically large number of dishes $N$, these requirements are not suitable for immediate quality control during ongoing production. Therefore, during production, we enforce that each individual feed parameter $p_i$ should fall within the accuracy threshold. Additionally, we require the dish surface measurement RMS error relative to its own best-fit paraboloid to remain strictly below $1.0$ mm.

\section{Dish Surface Accuracy and Precision} \label{sec:surface}
\begin{figure}[ht]
   \begin{center}
   \begin{tabular}{c} 
   \includegraphics{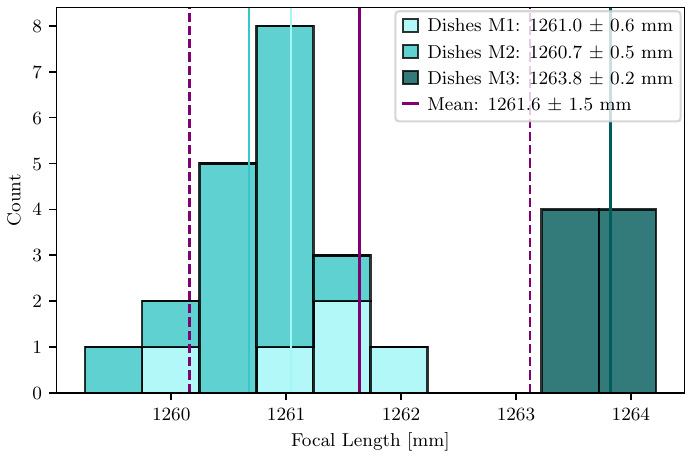} 
   \end{tabular}
   \end{center}
   \caption[example] 
   { \label{fig:fl_fac} 
A histogram of the best-fit focal lengths from factory measurements of the first 28 dishes produced for the Swartfontein site, grouped by their manufacturing molds. The ensemble mean best-fit focal length is indicated by the solid purple line, with dashed lines representing its standard deviation. The mean best-fit focal lengths for dishes from the same molds are indicated by their corresponding colors.}
\end{figure}

\begin{figure}[ht]
   \begin{center}
   \begin{tabular}{c c} 
   \includegraphics{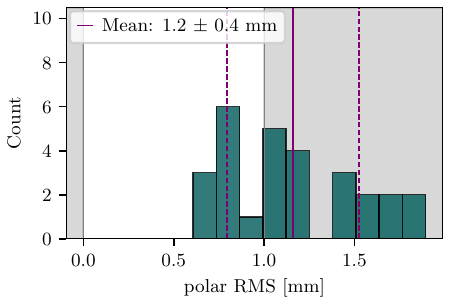} 
   & \includegraphics{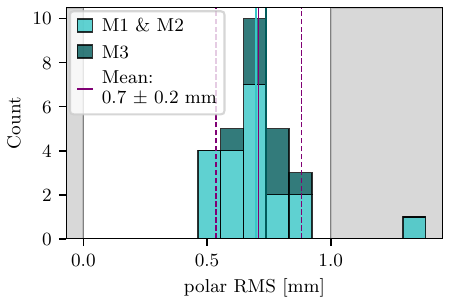}\\
   (a) & (b) \\
   \end{tabular}
   \end{center}
   \caption[example] 
   { \label{fig:prms_best-fit} 
Histograms showing the polar RMS of the dishes relative to the mean best-fit paraboloid. In Approach 1 (a), the mean best-fit is calculated across all dishes combined. In approach 2 (b), the dataset is segregated into dishes from molds 1 \& 2 and molds 3, with the mean best-fit calculated independently for each group.}
\end{figure}

The dishes are manufactured from fiberglass with an embedded aluminum mesh and are infused on a mold. Currently, three molds are in use, all of which were fabricated from a primary plug constructed out of wood and steel to form a precise paraboloid surface. The results of the molds and plug surface measurements are presented in~\cite{MetrologyST}. Each dish is supported on its underside by a rectangular structure, the so-called backing ring, which also serves as the mechanical interface to the mount as is visible in Figure \ref{fig:hirax}. The feed is held in place by four fiberglass legs attached to the dish.\\
All dishes are measured in the factory to ensure their surface deviations fall within our predefined tolerances as outlined in Section \ref{sec:requirements}.  The mean translational error of the points in the photogrammetry measurement dataset is found to be \qty{20}{\micro\meter}. For these measurements, the dishes are placed with their backing rings onto a surface well-leveled within \qty{1.5}{\arcmin}. The mounts are attached only after testing as they would raise the dishes too high for measurements to be performed in the production facility.\\
The best-fit focal length of each of the first 28 dishes produced for the Swartfontein site (MeerKAT National Park, South Africa) are shown in the histogram in Figure \ref{fig:fl_fac}. 
We find a mean best-fit focal length of  \qty{1261.6 +- 1.5}{\milli\meter} and the polar RMS for each dish remains below the required 1.0 mm threshold. The polar RMS is calculated by interpolating the measured data onto a polar grid and computing the root-mean square error from the residuals of the interpolated data relative to their best-fit paraboloids. The advantage of computing the RMS on a polar grid is that the polar RMS is independent of the spatial point distribution, unlike the RMS of the raw measurement set, which enables a more robust comparison across different dishes.\\
Figure \ref{fig:fl_fac} highlights that dishes fabricated from mold 3 (M3) have a mean focal length that differs by more than \qty{3}{\milli\meter} from those produced on molds 1 (M1) and 2 (M2). This discrepancy was expected due to a known degradation of the plug over time, which altered the focal length of mold 3 (discussed in further detail in~\cite{MetrologyST}{}). Because of this significant variance, we evaluate the surface accuracy and precision using two distinct approaches. In Approach 1, all dishes are treated as a single ensemble. In Approach 2, the dataset is segmented into two independent groups: dishes from molds 1 and 2, and dishes from mold 3.\\
Under Approach 1, the mean best-fit focal length is \qty{1261.6 +- 1.5}{\milli\meter}. Evaluating the mean best-fit paraboloid on a polar grid and comparing it to the fiducial paraboloid yields a maximum absolute residual of \qty{2.3}{\milli\meter}, comfortably satisfying the accuracy requirement. For the surface deviation precision the polar RMS is calculated per dish and plotted in Figure \ref{fig:prms_best-fit} (a). 
We find using Approach 1, 18 out of the 28 dishes exceed the 1.0 mm polar RMS threshold, meaning only 36\% of the dishes fulfill the precision requirement.\\
For Approach 2, the dataset yields two separate mean best-fit focal lengths: $1260.8 \pm 0.5$ mm for dishes from molds 1 and 2, and $1263.8 \pm 0.2$ mm for dishes from mold 3. As expected, sub-grouping yields a significantly smaller standard deviation than Approach 1. We find a surface accuracy of $1.1$ mm for molds 1 and 2 dishes, and $5.4$ mm for mold 3 dishes. Consequently, the accuracy requirement is only met for the dish group from molds 1 and 2. However, under Approach 2, only a single dish fails the precision threshold, corresponding to a 96\% compliance rate across the ensemble, as shown in Figure \ref{fig:prms_best-fit} (b).\\
The results of the dish surface analysis are summarized in Table \ref{tab:surface_results}. These results motivate evaluating a separate treatment for mold 3 dishes during calibration.
\begin{table}[h]
\centering
\begin{tabular}{|c|c|c|} 
\hline
Dish Surface & Accuracy (mm) & Precision Compliance (\%) \\
 \hline
Approach 1 & 2.3 & 36 \\
Approach 2 & 1.1 / 5.4 & 96 \\
\hline
\end{tabular}
\caption{The dish surface accuracy and precision is calculated using two different approaches. Approach 1 evaluates all dishes collectively, while Approach 2 analyzes dishes from molds 1 and 2 separately from those manufactured on mold 3.}
\label{tab:surface_results}
\end{table}

\subsection{Dishes at Site}
\begin{figure}[ht]
   \begin{center}
   \begin{tabular}{c} 
   \includegraphics{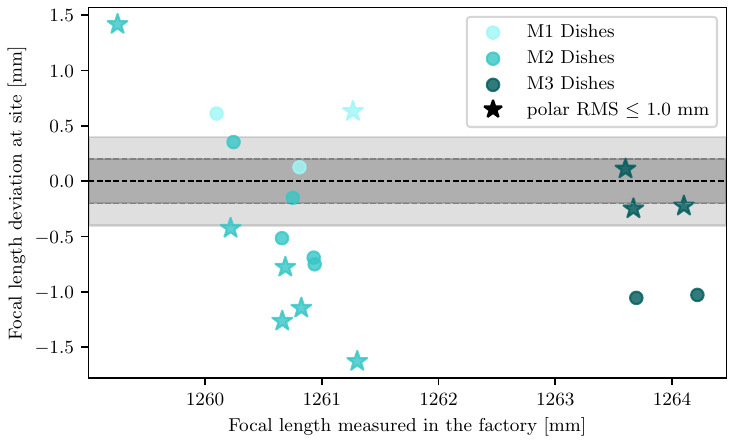} \\
   \end{tabular}
   \end{center}
   \caption[example] 
   { \label{fig:fl_site} 
The factory-measured best-fit focal lengths plotted against the corresponding on-site measured best-fit focal length deviations for the first 19 dishes, categorized by manufacturing molds. The dark gray and light gray shaded regions indicate the 1$\sigma$ and 2$\sigma$ measurement uncertainties of the best-fit focal length, respectively. This was determined by repeated measurements and surface analyses of a single dish.}
\end{figure}

The dish surfaces are also measured on-site using photogrammetry once they are installed on the mounts. Because the same measurement procedure is used, the spatial distribution of the measurement points remains consistent with the factory measurement data set. However, the mean translational measurement error of the photogrammetry system on-site (\qty{100}{\micro\meter}) is higher than that of the factory setup due to variable environmental factors, such as ambient lighting. Here, we analyze the dataset acquired in February 2026 for the first 19 dishes installed on-site in Swartfontein.\\ 
As discussed in~\cite{MetrologyST}{}, we identified quadrupole-like surface deviations induced by the mount due to mechanical stress during the initial on-site tests. Because these issues are currently being addressed and mitigated we forego analyzing the on-site polar RMS in this work, as it will still change significantly. However, it is insightful to determine whether the best-fit focal length is also affected by the mount and other effects on-site. The focal length determines the global surface accuracy of the dishes and establishes the nominal feed position used in Section \ref{sec:feed}.\\ 
We find a mean best-fit focal length from these on-site measurements of $1261.1 \pm 1.5$ mm. Figure \ref{fig:fl_site} illustrates the shift in the best-fit focal length measured on-site relative to the factory measurement result. 
The dark gray shaded area denotes the standard deviation $\sigma$, while the light gray region represents  the $2\sigma$ region on the best-fit focal length. This error was inferred from 25 repeated measurements of a single dish at the factory that remained stationary throughout.\\ 
The stars denote dishes whose on-site best-fit polar RMS remains below 1.0 mm, therefore, identifying dishes that are less dominated by the mount-induced issues. However, no clear trend is seen for dishes that are less affected by the mount-induced quadrupole distortions. Ultimately, these results demonstrate that on-site measurements are vital for an accurate characterization and understanding of the dish surface deviations post final assembly. Nevertheless, factory measurements provide a first approximation, particularly since the mean focal lengths are consistent with each other.

\section{Feed Placement Precision} \label{sec:feed}
\begin{figure}[ht]
   \begin{center}
   \begin{tabular}{c} 
   \includegraphics{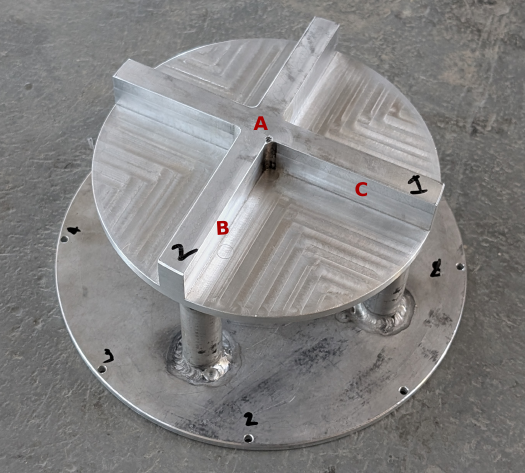} 
   \end{tabular}
   \end{center}
   \caption[example] 
   { \label{fig:verif_jig} 
The verification jig used to measure and verify the correct positioning of the feed.}
\end{figure}

\begin{figure}[ht]
   \begin{center}
   \begin{tabular}{c c} 
   \includegraphics{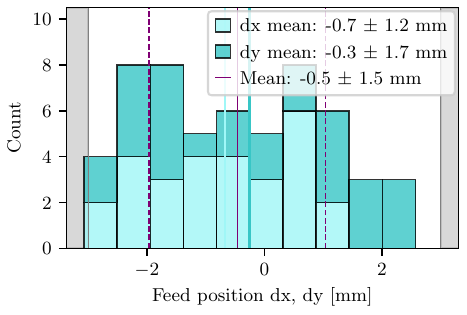} & \includegraphics{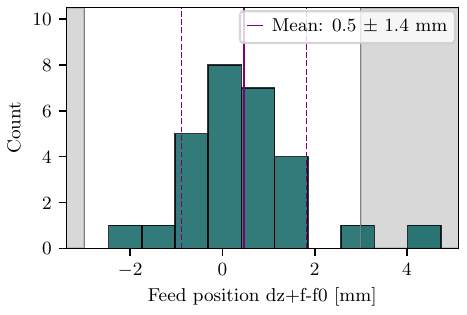}\\
   (a) & (b) \\
   \includegraphics{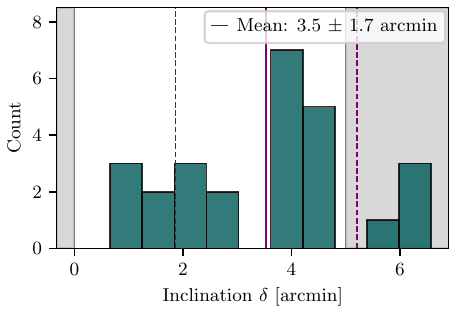} & \includegraphics{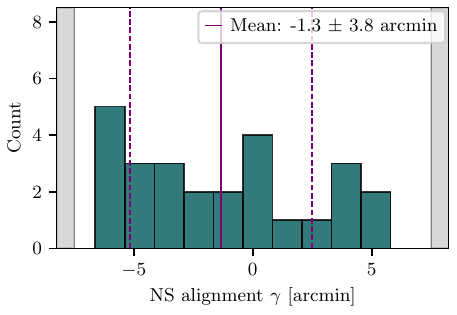}\\
   (c) & (d) \\
   \end{tabular}
   \end{center}
   \caption[example] 
   { \label{fig:feed_stat} 
Histograms of the five parameters characterizing the feed placement. The spatial offset from the nominal focal point in $x$ and $y$ direction are given by $\mathrm{d}x$ and $\mathrm{d}y$ in (a). The offset along the boresight axis relative to the fiducial position for a fiducial focal length dish is shown in (b) as $\mathrm{d}z+f-f_0$. The inclinations and the alignment offsets are shown in (c) and (d), respectively.}
\end{figure}

The feed placement is measured using photogrammetry with a dedicated, precision-machined, verification jig (shown in Figure \ref{fig:verif_jig}) mounted in the same way as the feed and can. 
Photogrammetry stickers are attached to surfaces A, B, and C marked in Figure \ref{fig:verif_jig} and are used to determine the center point of the feed, given by the intersection of these three planes. The declination $\delta$ is given by the inclination of surface A with respect to the coordinate system of the best-fit paraboloid. The N-S alignment $\gamma$ is determined from the offset of the plane of surface B from the N-S line marked on the dish with photogrammetry stickers. This measurement procedure can only be performed in the factory prior to commissioning of the feeds, meaning that all resulting values are referenced to the best-fit paraboloid of each dish as measured in the factory.\\
Figure \ref{fig:feed_stat} displays histograms of the feed parameters measured across the first 28 dishes. The ensemble mean values (solid line) and their corresponding standard deviations (dashed line) are marked by vertical lines. The white windows mark the allowable accuracy boundaries. Recall from Section \ref{sec:requirements} that the accuracy requirement is defined with respect to the mean value of the parameter. As shown in Figure \ref{fig:feed_stat}, the accuracy is met successfully for all parameters.\\
The achieved accuracy and precision for the feed parameters are summarized in Table \ref{tab:feed_accprec}. 
For the set of dishes included in this analysis, only the feed inclination angle $\delta$ strictly satisfies the target precision requirement. The parameters $\mathrm{d}x$, $\mathrm{d}y$ and $\gamma$ scatter significantly and are distributed almost uniformly across the non-shaded allowable regions leading to a high precision value above the threshold. For $\gamma$ in particular, data points are noticeably clustered near the accuracy threshold boundary. This distribution arises from the fact that the photogrammetry equipment is utilized not only for quality control, but also as an active feedback mechanism during production. If an assembly measurement falls outside the accuracy requirement, the positioning is manually adjusted and the dish is remeasured. This process is repeated until the installation passes the quality control and therefore leads to a higher precision value.\\ 
The height of the feed within the dish is fixed with a pole that is inserted into the center drainage hole of the dish during assembly, enabling a precise vertical placement provided the pole is well oriented. We use $\mathrm{d}z+f-f_0$ to quantify the precise and accurate placement of the feed height with respect to the fiducial focal length. The outliers observed in $\mathrm{d}z+f-f_0$ occurred because originally $\mathrm{d}z$, which is the position relative to the individual best-fit focal length, was used as the target for the positional requirement along this axis. However, due to the different focal length of mold 3, this strategy would require the use of two poles with different lengths for the installation of the feed in dishes from different molds. To ensure manufacturing repeatability, we opted to  prioritize a consistent absolute $z$-height over forcing each dish perfectly into focus. The two dishes falling within the gray shaded area are mold 3 dishes produced before this shift to $\mathrm{d}z+f-f_0$ was implemented; these can be corrected with mechanical spacers on-site if deemed necessary. Aside from these outliers, the $\mathrm{d}z+f-f_0$ distribution demonstrates that the pole enables a repeatable vertical positioning. While the protocols for installing $\mathrm{d}x$, $\mathrm{d}y$, and $\gamma$ repeatably are still being refined, tracking these metrics will allow us to steadily drive these precision values towards our targets as the number of dishes increases.
\begin{table}[t]
\centering
\begin{tabular}{|c|c|c|c|} 
\hline
Feed Parameters & Achieved [Target] Accuracy & Achieved [Target] Precision & Units \\
 \hline
Feed position $\mathrm{d}x$ & 0.7 [3.0] & 1.2 [0.5] & mm\\
Feed position $\mathrm{d}y$ & 0.3 [3.0] & 1.7 [0.5] & mm\\
Feed position $\mathrm{d}z+f-f_0$ & 0.5 [3.0] & 1.3 [0.5] & mm\\
Feed inclination angle $\delta$ & 3.5 [5.0] & 1.6 [2.5] & arcmin \\
N-S alignment $\gamma$ & 1.3 [7.5] & 3.8 [2.5] & arcmin \\
\hline
\end{tabular}
\caption{The accuracy and precision values achieved for the feed parameters directly compared to the target threshold from the requirements.}
\label{tab:feed_accprec}
\end{table}

\section{Conclusions}
This paper presents a statistical analysis of the measured dish surfaces and feed positioning of the first 28 HIRAX dishes, comparing them against the requirements established in the HIRAX Telescope Mechanical Assembly Requirements Document. From factory measurements, we found a mean focal length of $1261.6 \pm 1.5$ mm. However, because the focal lengths of mold 3 dishes systematically differ by approximately $3$ mm from the other dishes, we perform the analysis of the dataset using two distinct analytical frameworks. In Approach 1, the dishes were treated as a single ensemble; in Approach 2, the mold 3 dishes were separately treated from the other dishes. While Approach 1 satisfies the surface accuracy requirements, only 36\% of the dishes meet the precision requirements. Under Approach 2, 96\% of the dishes achieve the precision threshold, though the accuracy targets are only met by the mold 1 and 2 dish group. Initial on-site measurements underscore the necessity of post-assembly metrology for accurate surface characterization, provided that the current mount-induced stresses are successfully mitigated.\\
An evaluation of the feed placement parameters found that the accuracy requirements of all the parameters are met. However, only the feed inclination angle $\delta$ meets the target precision threshold, whereas the translational offsets and the N-S alignment $\gamma$ display significant scatter. This highlights that current manual feed deployment protocols require further engineering refinement to ensure repeatability and achieve the required precision.\\
In the next phase of this work, we will propagate these measured physical deviations into simulated primary beams. These perturbed beams will be used to model the exact effect of these instrumental systematic errors on the cosmological signal. Ultimately, these simulations will determine whether Approach 1 or 2 provides a more robust calibration for HIRAX's redundant array. Opting for Approach 2 implies that redundant calibration cannot treat the array as a single, uniform population, but must instead accommodate two distinct sub-ensembles with different primary beam responses. This is a vital step towards suppressing systematics-induced foreground contamination and enabling precision cosmological constraints with HIRAX's intensity mapping survey.

\acknowledgments 
The authors would like to thank Advanced Fibreform\footnote{\url{https://carbon-fibre.co.za/}} and their staff for the dish production and all their provided support. We also want to thank Hexagon for their excellent support provided for the photogrammetry system that was purchased through them. We thank Siyabonga Ngcobo for his support with some of the measurements, and Nicola Lo Russo for his project management support for the Swiss HIRAX team.\\
This work was supported in part by the SNSF Flare Grant 216653, the SNSF Lead Agency grant 10.001.432, and the SNSF project grant 200021\_192243. KM acknowledges support from the National Research Foundation of South Africa. H.C.C. acknowledges the support of an NSERC Discovery (RGPIN-2019-04506) and NSERC Alliance International grant (ALLRP 586202-23). WN acknowledges the financial assistance of the South African Radio Astronomy Observatory (SARAO) towards this research (\url{www.sarao.ac.za})\\
HIRAX is hosted by the South African Radio Astronomy Observatory (SARAO), a facility of the National Research Foundation (NRF). We are grateful to the SARAO technical and site personnel for their support in deployment, commissioning and on-site operations.\\
This work made use of the astropy~\cite{astropy:2022, astropy:2018, astropy:2013}{}, healpy~\cite{Zonca2019}{} from HEALPix~\cite{Healpix2005}, numpy~\cite{Harris2020}, scipy~\cite{Virtanen2020}{} and matplotlib~\cite{Hunter2007} software packages.

\bibliography{report} 
\bibliographystyle{spiebib}

\appendix    
\section{Dish Surface Accuracy and Precision by Individual Molds}
\begin{figure}[ht]
   \begin{center}
   \begin{tabular}{c} 
   \includegraphics[width=0.48\textwidth]{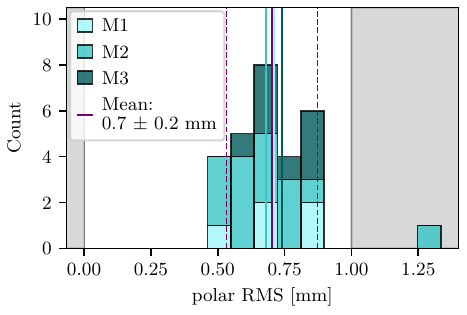} 
   \end{tabular}
   \end{center}
   \caption[example] 
   { \label{fig:prms_meanM1M2M3} 
A histograms of the polar RMS of the dish surfaces with respect to the mean best-fit paraboloid calculated independently for dishes from individual manufacturing molds.}
\end{figure}

We extend the analysis described in Section \ref{sec:surface} by segregating the dishes into three distinct groups corresponding to the molds on which they were fabricated. This approach yields the following best-fit focal lengths: $1261.0 \pm 0.6$ mm for dishes from Mold 1, $1260.7 \pm 0.5$ mm for dishes from Mold 2 and $1263.8 \pm 0.2$ mm for dishes from Mold 3. These groups result in individual surface accuracies of $1.5$ mm, $1.0$ mm and $5.4$ mm, respectively.\\ 
Reflecting the results of Approach 2, the mold 3 dish group fails the accuracy requirement because its configuration is identical to that used in the previous approach. The calculated precision closely matches the results of Approach 2, where 96\% of the dishes successfully met the target precision as illustrated in Figure \ref{fig:prms_meanM1M2M3}.

\end{document}